\documentclass[9,preprint,superscriptaddress,amsmath,amssymb,prl,floatfix,a4paper,onecolumn]{revtex4-1} 
\usepackage[ngerman,english]{babel}
\usepackage{booktabs,setspace}
\usepackage{xspace}
\usepackage{amsmath,tensor}
\usepackage{array,color}
\usepackage{relsize,exscale}  
\usepackage{textcase}	
\usepackage{soul,color}		
\usepackage{calc} 
\usepackage{makeidx,showidx}
\usepackage{graphicx}
\usepackage{booktabs}
\usepackage{wrapfig}
\usepackage{float}
\usepackage{textcomp}
\usepackage{siunitx}
\usepackage[version=3]{mhchem}
\usepackage[mathlines]{lineno}

\newcommand{\lsco} {{La$_{2-x}$Sr$_x$CuO$_4$}\@\xspace}

\newcommand{\hgryb} {{HgBa$_{2}$CuO$_{4+\delta}$}\@\xspace}

\newcommand{\ybco} {{YBa$_2$Cu$_3$O$_{7-\delta}$}\@\xspace}

\newcommand{\ybcoE} {$\ce{YBa2Cu4O8}$\@\xspace}

\newcommand{\tc} {\ensuremath{T_{\mathrm c}}\@\xspace}

\begin{document}
\title{Temperature independent cuprate pseudogap from planar oxygen NMR}

\author{Jakob Nachtigal}
\author{Marija Avramovska}
\affiliation{Felix Bloch Institute for Solid State Physics,
 Leipzig University, Linn\'estraße 5, 04103 Leipzig, Germany}
 
\author{Andreas Erb}
\affiliation{Walther Meissner Institut, Bayerische Akademie der Wissenschaften, 85748 Garching, Germany}

\author{Danica Pavi\'cevi\'c}
\author{Robin Guehne}
\author{J\"urgen Haase}
\affiliation{Felix Bloch Institute for Solid State Physics,
 Leipzig University, Linn\'estraße 5, 04103 Leipzig, Germany}



\date{\today}




\begin{abstract}
Planar oxygen nuclear magnetic resonance (NMR) relaxation and shift data from all cuprate superconductors available in the literature are analyzed. They reveal a temperature independent pseudogap at the Fermi surface, which increases with decreasing doping in family specific ways, i.e., for some materials the pseudogap is substantial at optimal doping while for others it is nearly closed at optimal doping. The states above the pseudogap, or in its absence are similar for all cuprates and doping levels, and Fermi liquid-like.  If the pseudogap is assumed exponential it can be as large as about 1500 K for the most underdoped systems, relating it to the exchange coupling. The pseudogap can vary substantially throughout a material, being the cause of cuprate inhomogeneity in terms of charge and spin, and consequences for the NMR analyses are discussed. This pseudogap appears to be in agreement with the specific heat data measured for the YBaCuO family of materials, long ago. Nuclear relaxation and shift show deviations from this scenario near $T_{\mathrm{c}}$, possibly due to other in-gap states.
\end{abstract}


\maketitle



\section{Introduction}
Nuclear magnetic resonance (NMR) provides important local information about the electronic properties of materials \cite{Slichter1990}, and it has played a key role in the characterization of cuprate high-temperature superconductors \cite{Slichter2007, Walstedt2008}. However, different from when NMR proved BCS theory \cite{Bardeen1957, Hebel1957}, for cuprates a full theoretical understanding is lacking, and thus, it is challenging to decipher NMR data. 

In classical metals and superconductors, NMR is known for the local measurement of the electronic spin susceptibility \cite{Heitler1936,Knight1949,Korringa1950,Schumacher1956,Yosida1958}, including the predicted changes in the density of states at the Fermi surface with a coherence peak in nuclear relaxation \cite{Hebel1957}. In the normal state, the high density of states near the Fermi surface leads to the distinctive, fast nuclear relaxation ($1/T_1$) that is proportional to temperature ($1/T_1 \propto T$) since temperature increases the available number of electronic states for scattering with nuclear spins. Quite to the contrary, the NMR spin shift that is proportional to the uniform electronic spin susceptibility is temperature independent, as the increase in temperature also decreases the occupation difference. 

These elements of observation were the backdrop against which the cuprate NMR data were discussed, early on. Unfortunately, the cuprates have large unit cells and the important nuclei in the plane, $^{63,65}$Cu and $^{17}$O, have electric quadrupole moments and thus are affected by the local charges, as well. This leads to multiple resonances that have to be assigned to the chemical structure, and inhomogeneously broadened lines in the non-stoichiometric systems are the rule. This complicates measurement and interpretation. Fortunately, the cuprates are type-II materials and can be investigated in the mixed state below \tc at typical magnetic fields used for NMR, which gives access to the properties of the superfluid, but also complicates shift measurements from residual diamagnetism \cite{Barret1990b}.

Early on, a number of more or less universal magnetic properties of the cuprates were derived, such as spin-singlet pairing, the pseudogap, and special spin fluctuations (for reviews see \cite{Slichter2007,Walstedt2008}). Here, we will not dwell on a more detailed discussion of previous conclusions, as we believe that while the data are undisputed, the prevailing view needs to be corrected. 

In recent years, some of us were involved in special NMR shift experiments that raised suspicions about the description of the magnetic properties based on NMR \cite{Haase2009b,Meissner2011,Haase2012,Rybicki2015}. During the same period of time, a comprehensive picture of the charge distribution in the CuO$_2$ plane was developed \cite{Zheng1993,Haase2004,Jurkutat2014}. It fostered the understanding of charge sharing in electron and hole doped cuprates, as it was found that $1+x = n_{\mathrm{Cu}} + 2 n_{\mathrm{O}}$, i.e., the charges measured with NMR in the planar Cu ($n_{\mathrm{Cu}}$) and O ($n_{\mathrm{O}}$) bonding orbitals add up to the total charge, inherent plus doped hole ($x>0$) or electron ($x<0$) content. An astonishing correlation appeared in this context, as the maximum \tc of a curpate system ($T_{\mathrm{c,max}}$) is nearly proportional to $n_{\mathrm{O}}$ \cite{Jurkutat2014,Rybicki2016}. This explains the differences in $T_{\mathrm{c,max}}$ between the various families that differ in charge sharing considerably, and it calls into question the usefulness of what one calls the cuprate phase diagram, rather, a phase diagram in terms of $n_{\mathrm{Cu}}$ and $n_{\mathrm{O}}$ appears advantageous \cite{Jurkutat2019b}.

These findings suggested that some cuprate properties might be family dependent, and that a broader look at NMR data might be useful, as well. 
Since planar O NMR requires the exchange of $^{16}$O by $^{17}$O, which is not easily performed for single crystals and can have consequences for the actual doping and its spatial distribution, the focus was on planar Cu data that appeared more abundant and more reliable. 

Immediately, the overview of the Cu shifts across all families \cite{Haase2017} demands different shift and hyperfine scenarios, as the changes in the shifts are not proportional to each other (similar to what was found with special NMR experiments before \cite{Haase2009b,Haase2012,Rybicki2015}). Likely, it involves two spin components, one that has a negative uniform response and is located at planar Cu, coupled to a second component (presumably on planar O) with the usual positive response. 
In a next step, all planar Cu relaxation data were gathered \cite{Avramovska2019,Jurkutat2019}, and from the associated plots it became obvious that, surprisingly, the Cu relaxation is quite ubiquitous, very different from what was concluded early on. It turns out that the relaxation rate measured with the magnetic field in the plane ($1/T_{1\perp}$) does neither change significantly between families, nor as a function of doping, with $1/T_{1\perp}T_{\mathrm{c}}\approx 21$/Ks. Only the relaxation anisotropy changes by about a factor of three across all cuprates. Thus, no enhanced, special spin fluctuations are present in the underdoped systems. This leaves as an explanation for the failure of the Korringa relaxation (discovered early on \cite{Walstedt2008}) only a suppression of the NMR shifts \cite{Avramovska2019}. 
This also means that there is no pseudogap effect in planar Cu relaxation, while the Cu shifts do have a temperature dependence above \tc presumably from pseudogap effects. Finally, it was shown that the planar Cu relaxation can be understood in terms of two spin components, as well \cite{Avramovska2020}, where a doping dependent correlation of the Cu spin with that of O explains the relaxation anisotropy. Furthermore, the unusual planar Cu shift component that is a function of doping and not necessarily temperature was found to be present in the planar O high temperature data \cite{Pavicevic2020}, where it causes the hallmark asymmetry of the total quadrupole lineshape, observed long ago \cite{Takigawa1989b,Kambe1993,Haase2000}, but not understood.

\begin{figure}[t]
\centering
\includegraphics[width=0.9\textwidth ]{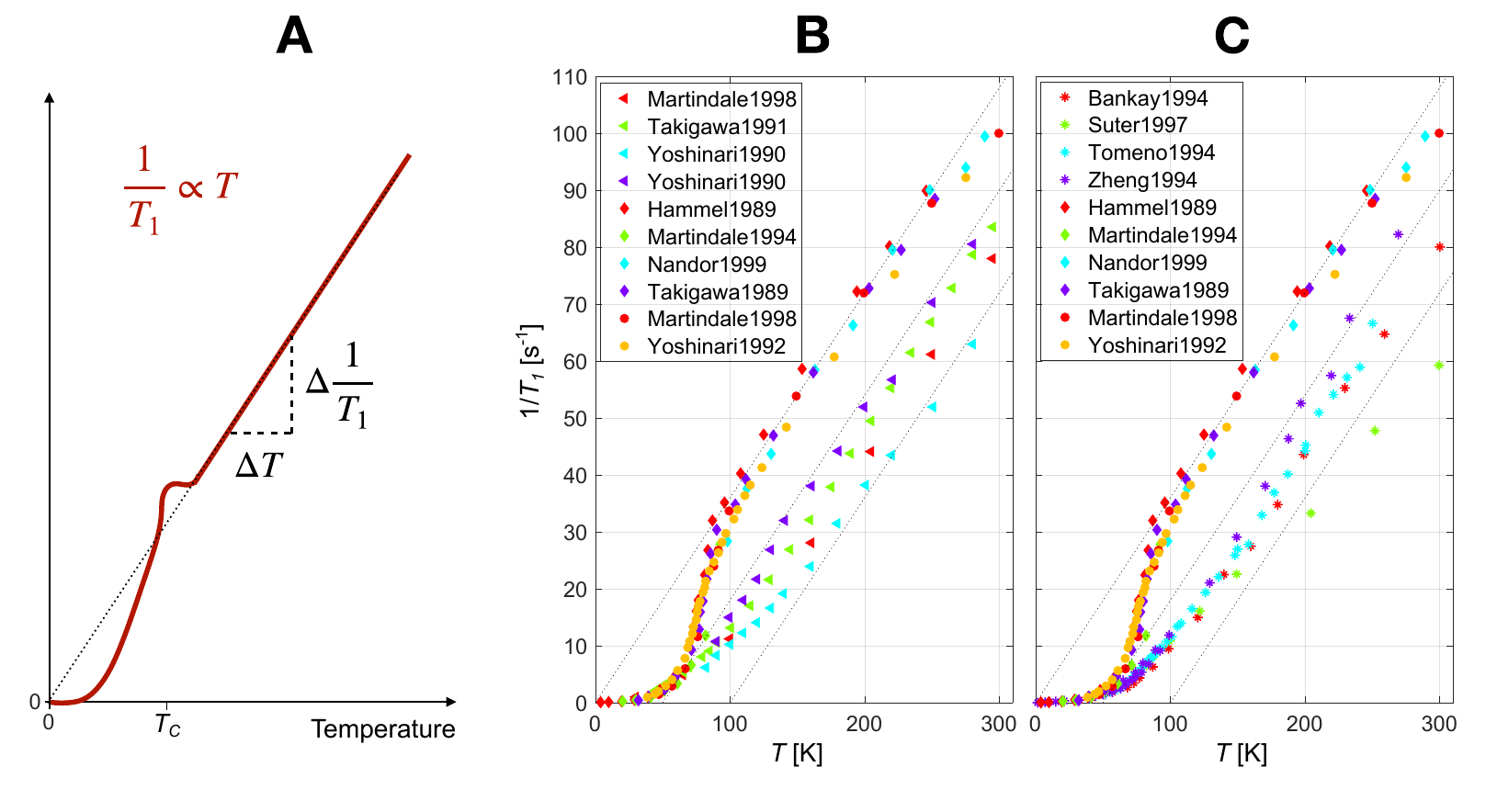}
\caption{Nuclear Relaxation. {\bf A} (sketch), above the critical temperature for superconductivity, $T_{\mathrm{c}}$, in a Fermi liquid, the  relaxation is proportional to temperature, i.e., the slope points to the origin of the plot; only just below $T_{\mathrm{c}}$, the BCS gap for spin singlet pairing leads to a loss of states and relaxation (after the Hebel-Slichter coherence peak). {\bf B}, optimally doped YBa$_2$Cu$_3$O$_{6.96}$ (full circles) and overdoped YBa$_2$Cu$_3$O$_7$ (diamonds) behave Fermi liquid-like above $T_{\mathrm{c}}$ (dotted lines have slope $1/(T_1T)$=0.36{/Ks}). Underdoped YBa$_2$Cu$_3$O$_{7-\delta}$ (triangles) show identical high-temperature behavior in the sense that as a function of temperature the relaxation increases with the same slope as found for optimally and underdoped systems, i.e., as the Fermi function opens with increasing temperature, it adds states at the same rate. However, the slope does not intersect the origin, which shows that even  at high temperatures, low energy states are missing. This is the planar O pseudogap effect that rapidly evolves when the doping is lowered. {\bf C}, same as {\bf B}, except the relaxation data for the underdoped materials have been replaced by data for YBa$_2$Cu$_4$O$_8$ (starred points); this underdoped, stoichiometric material displays a very similar temperature dependence at higher temperatures. For the references see Appendix {\bf A}.           
}\label{fig:fig1}
\end{figure}

Here, we present all temperature dependent shift and relaxation data of planar $^{17}$O collected in an intensive literature search (data points from about 80 publications were taken). 
The main conclusion from the data will be that planar O relaxation, different from Cu, is affected by the pseudogap that also dominates the planar O shifts. Here, the pseudogap represents itself as a loss in density of states close to the lowest energies (at the Fermi surface) for the underdoped materials, and this gap is temperature independent, but set by doping, different from what is often assumed \cite{Mukhopadhyay2019,Sato2017}. This scenario is in agreement with early specific heat data \cite{Loram1998} that also discussed such a pseudogap in \ybco. The largest found pseudogap is in agreement with a node-less suppression of states of the size of the exchange coupling, \SI{1500}{K}. It rapidly decreases with increasing doping, e.g., it is closed for \ybco at optimal doping, but not for optimally doped \lsco.

\vspace{0.4cm}
\section{Planar Oxygen Relaxation and Shift for YBa$_2$Cu$_3$O$_{6+y}$ and YBa$_2$Cu$_4$O$_8$}
Nuclear relaxation of planar oxygen shows strikingly simple behavior in these most studied materials, and we will find the conclusions to be generic to the cuprates.
 
\subsection{Planar Oxygen Relaxation}
In Fig.~\ref{fig:fig1}, next  to a sketch of expected behavior for a Fermi liquid ({\bf A}) we plot the relaxation rate ($1/T_1$) vs. temperature ($T$). It is apparent that optimally and overdoped \ybco ({\bf B}) are Femi liquid-like: above \tc, an increase (decrease) in temperature adds (subtracts) additional states for nuclear scattering and even the density of states (DOS) seems to be rather constant up to about \SI{250}{K} (above that temperature the relaxation appears to begin to lag behind the expected value \cite{Nandor1999}). 

It is important to note that at high temperatures, changes in temperature ($\Delta T$) lead to proportional changes in relaxation ($\Delta (1/T_1)$) with a slope of \SI{0.36}{/Ks} that intersects the origin. With other words, the proportionality of the rate to temperature is only disturbed by the opening of the superconducting gap at \tc, below which relaxation drops more rapidly as pairing sets in (no Hebel-Slichter peak is observed). Thus, planar O relaxation of optimally and overdoped \ybco appears determined by Fermi liquid-like electrons, turning into a spin singlet superconductor.

The underdoped materials behave distinctively different, Fig.~\ref{fig:fig1}{\bf B}. Here we observe a rapid change of relaxation with doping at given temperature, but we find nearly the same high-temperature slope of about \SI{0.36}{/Ks}, i.e., increasing the temperature adds states at the same rate as for optimally or overdoped systems. However, the shifted slopes signal an offset in temperature below which relaxation must disappear. This means, even at much larger temperatures one is aware of the lost, low temperature states. This is exactly what one expects if a temperature independent, low-energy gap in the DOS develops with doping (a gap that remains open at high temperatures). The same scenario applies to YBa$_2$Cu$_4$O$_8$, cf.~Fig.~\ref{fig:fig1}{\bf C}, where the intercept of the high-temperature slope with the abscissa is about \SI{70}{K}.

\begin{figure}[t]
\centering
\includegraphics[width=0.9\textwidth ]{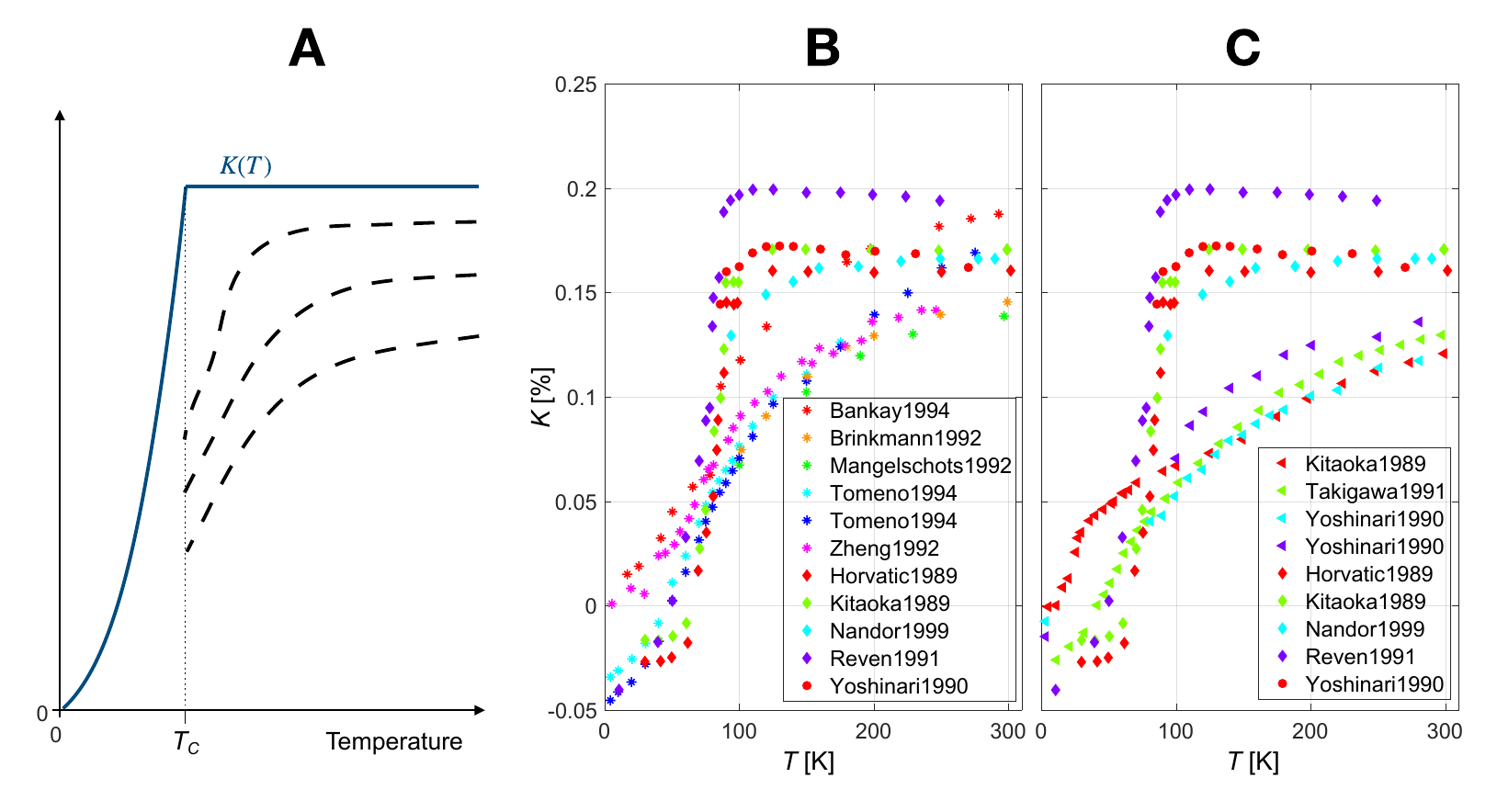}
\caption{Planar $^{17}$O NMR Shifts. {\bf A}, (sketch) Fermi liquid behavior with spin singlet pairing at $T_{\mathrm{c}}$ is shown with the full blue line. The dashed lines indicate what one expects based on the relaxation data: above $T_{\mathrm{c}}$, states are missing increasingly as the doping decreases, and as a function of temperature these lost states become more pronounced. {\bf B, C}, literature shift data. Optimally doped YBa$_2$Cu$_3$O$_{6.96}$ (circles) and overdoped YBa$_2$Cu$_3$O$_7$ (diamonds) behave Fermi liquid-like, but the underdoped materials YBa$_2$Cu$_3$O$_{7-\delta}$ (triangles), and YBa$_2$Cu$_4$O$_8$ (stars) show the expected high-temperature behavior. Below $T_{\mathrm{c}}$, the shifts drop less dramatically for the underdoped systems. Some materials appear to show a negative spin shift at the lowest temperatures. For the references see Appendix {\bf A}.         
}\label{fig:fig2}
\end{figure}

At lower temperatures, the rates  for \ybco become rather doping independent, below about \SI{80}{K}. It appears that the  special temperature dependence due to the superconducting gap and pseudogap merge, somewhat different from the behavior with \ybcoE, but still similar in the sense that the relaxation begins to increase as it departs from the parallel lines.

Note, the relaxation ceases completely at the lowest temperatures for all materials. While electric contributions (electric quadrupole interaction) to the relaxation have been shown to exist and contribute at lower temperatures \cite{Suter1997,Takigawa1991} their contribution vanishes, as well. The true magnetic relaxation dependences might be systematically shifted to lower rates at lower temperatures compared to what is seen in Fig.~\ref{fig:fig1}. Therefore, the apparent increase in relaxation could signal quadrupolar relaxation, as well. A thorough study of these effects might be in order.

\subsection{Planar Oxygen Shifts}
For planar O the orbital shift is almost negligible \cite{Takigawa1989b}, making the spin shifts rather reliable with uncertainties arising only from the diamagnetic response below \tc. Shift referencing is simple, as well, as ordinary tap water can be used for $^{17}$O NMR referencing (there is significant confusion about Cu shift referencing in the literature \cite{Haase2017}). 
Nevertheless, there appear to be deviations between the shifts measured on similar samples, even for stoichiometric \ybcoE \cite{Brinkmann1998}, and it is not always clear if shifts were corrected for the diamagnetic response. We will show the bare shifts without correction, in order to avoid introducing systematic errors. For example, it is possible that the uniform spin response from Cu$^{2+}$ is negative \cite{Haase2017,Avramovska2019} leading to a negative term for planar O at low temperatures.

Note that the diamagnetic response of the cuprates was experimentally determined with $^{89}$Y NMR, early on \cite{Barret1990b}, by assuming that this nucleus' spin shift is negligible at low temperatures ($4.2$K). A value of about 0.05\% was derived \cite{Barret1990b}. This value appears to be rather large \cite{Oldfield1989}, and as experiments with $^{199}$Hg NMR of \hgryb showed \cite{Rybicki2015}, the diamagnetic response measured at $^{199}$Hg is probably less than 0.01\% (note that $^{199}$Hg is located far from the plane and should not suffer from large spin shifts, different from $^{89}$Y that might be affected by a negative term, as well). 

For a Fermi liquid with a fixed DOS near the Fermi surface one expects a temperature independent spin shift ($K$) above \tc, since an increase in temperature adds new states from an opening Fermi function, but the occupation decreases at the same rate, cf.~Fig.~\ref{fig:fig2}{\bf A}. Now, in view of the planar O relaxation, a temperature independent gap at the Fermi surface should be assumed. Then, qualitatively, we expect a behavior shown in Fig.~\ref{fig:fig2}{\bf A}: at the highest temperatures, far above the gap, low temperature states will still be missing, leading to a lower spin shift. As the temperature is lowered, the effect of the gap will be more severe. This is in agreement with data in Fig.~\ref{fig:fig2}{\bf B} and {\bf C}. Below \tc, we note that there is no sudden loss of states as for optimally or overdoped materials, which one might naively expect if the same superconducting gap opens on the states still available. Quite to the opposite, a less rapid decrease of the shifts below \tc is observed (we noted a different low temperature behavior for relaxation, as well).

Note that the Korringa relation is given by $T_1TK^2=(\gamma_e/\gamma_n)^2\hbar/(4\pi k_B)\equiv S_0$ \cite{Korringa1950}, and with $S_0 = 1.4 \cdot 10^{-5}$ Ks one estimates a spin shift of about $K=0.23$\% from the relaxation slope of \SI{0.36}{/Ks}, not very different from what is observed for optimally or overdoped systems in Fig.~\ref{fig:fig2}.

\subsection{Numerical Analysis}

\begin{figure}[t]
\centering
\includegraphics[width=1\textwidth ]{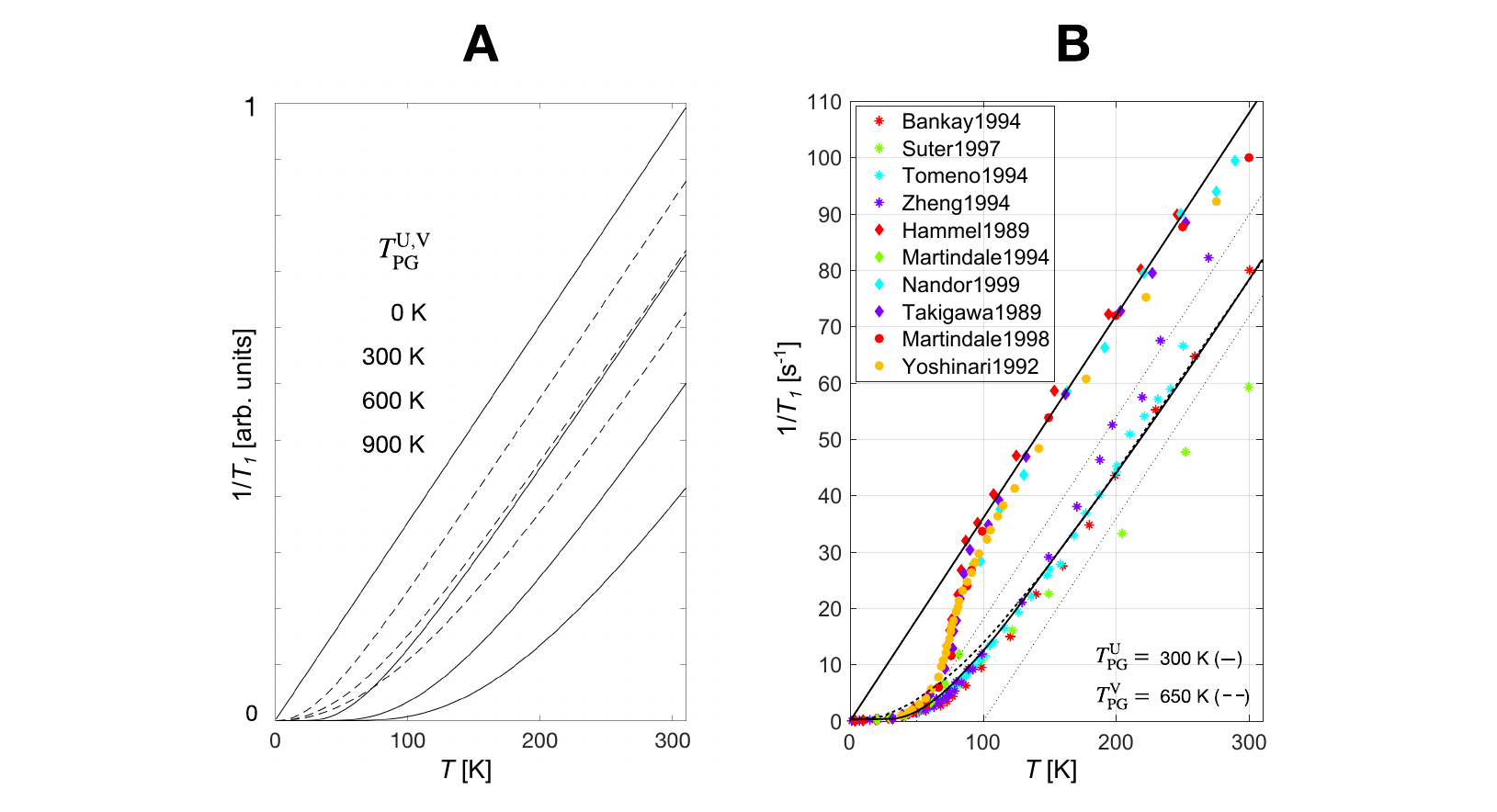}
\caption{ {\bf A}, model relaxation calculations with a U- and V-shaped gap ($T^{\mathrm{U,V}}_{\mathrm{PG}}$) in the density of states. {\bf B}, estimation of the pseudogap temperature by varying the gap size for YBa$_2$Cu$_4$O$_8$.
}\label{fig:fig3}
\end{figure}

The planar O relaxation data point to a pseudogap that is simply caused by missing low energy states. This gap is not temperature dependent, but rapidly increases with decreasing doping. In a very simple picture (that is very likely \emph{not} to be correct, already in view of the planar Cu shift and relaxation data \cite{Haase2017,Avramovska2019,Jurkutat2019,Avramovska2019}), we use the Fermi function with fixed DOS and calculate the relaxation as being proportional to the sum of the product of occupied states times empty states (the nuclear energy change is negligible for the electrons), i.e. $\sum_E{p(E) [1-p(E)]}$, where 
\begin{equation}\label{eq:one}
p(E,\mu)=1/\left[1+\exp(E-\mu)/k_BT\right].
\end{equation}
As a result one finds the Heitler-Teller dependence \cite{Heitler1936}, $1/T_1 \propto T$, cf. Fig.~\ref{fig:fig3}. 

Now, one can remove manually states near the Fermi surface with a width $\Delta E$ given in temperature as defined by,
\begin{equation}\label{eq:two}
T_{\mathrm{PG}}^{\mathrm{U,V}} = \Delta E^{\mathrm{U,V}}/k_B,
\end{equation}
by assuming a U- or V-shaped gap in the DOS, respectively \cite{Loram1998}. For the U-shaped gap all states within $\Delta E$ are removed (exponential decrease), for a V-shaped gap a linear decrease in DOS is assumed, vanishing at $E=\mu$. This simple scenario leads to the found behavior, i.e. we obtain nearly parallel high-temperature lines for different sizes of this pseudogap, cf.~Fig.~\ref{fig:fig3}{\bf A}. For a given offset, the cutoff temperature is different for both gaps, cf.~Fig.~\ref{fig:fig3}{\bf B}.  With such an approach we find for \ybcoE a gap of about $T_{\mathrm{PG}}^{\mathrm{U}}\approx 300$ K ($T_{\mathrm{PG}}^{\mathrm{V}}\approx 650$ K). Obviously, one cannot decide on the shape of the gap. Note that the BCS gap is not included in the fit and that there are uncertainties from quadrupolar relaxation at lower temperatures. 

Since the action of the gap is to cause a near parallel shift of the high-temperature dependence, any spatial inhomogeneity of the gap will lead to similar lines, as well, very different from how it affects the shifts that we will discuss now. 

\begin{figure}[t]
\centering
\includegraphics[width=.9\textwidth ]{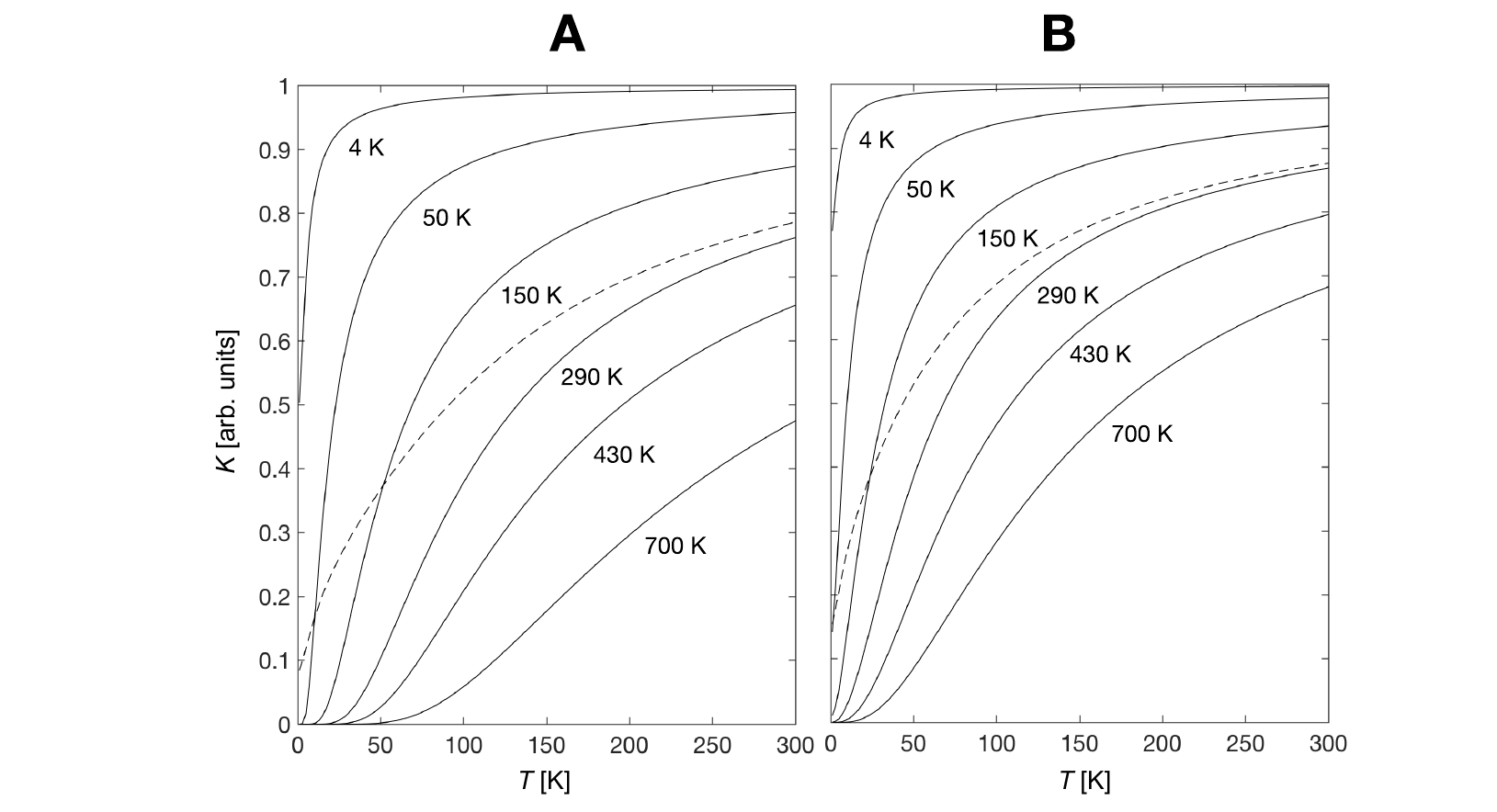}
\caption{Model calculations of temperature dependent shifts from a pseudogap at the Fermi surface with the indicated gap temperatures. {\bf A}, for a U-shaped gap, and, {\bf B} for a V-shaped gap. The simple mean of the shifts is indicated by a dashed line, emphasizing that a gap inhomogeneity can cause a different temperature dependence of the apparent magnetic shift. Also, the magnetic linewidths will behave differently (the linewidths will grow as the temperature decreases, before it finally decreases).         
}\label{fig:fig4}
\end{figure}

One can estimate what such a pseudogap will do for the NMR shifts (by assuming a slightly different $\mu$ for spin up and down). Examples are shown in Fig.~\ref{fig:fig4} for various $T_{\mathrm{PG}}^{\mathrm{U}}$ ({\bf A}), and $T_{\mathrm{PG}}^{\mathrm{V}}$ ({\bf B}). Clearly, for small gap sizes the shift will approach the Fermi liquid value (normalized to 1). The V-shaped gap has more total DOS and the action of the gap is weaker. 

Above \tc, one should be able to fit the experimental shifts, and by comparing Figs.~\ref{fig:fig4} and \ref{fig:fig2} one finds qualitative agreement. However, a more quantitative determination of the gap appears difficult since (i) there is a large spread in shifts already for similar samples, and (ii) at lower temperatures the shifts for the underdoped systems appear larger, cf. Fig.~\ref{fig:fig2}, pointing to gap inhomogeneity. Note that the dashed lines in Fig.~\ref{fig:fig4} are the simple mean shift of the shown temperature dependences. Thus, any spatial distribution of the pseudogap will change the actual temperature dependence as smaller gaps will lift the apparent shift at lower temperatures. We estimate gap sizes of $T_{\mathrm{PG}}^{\mathrm{U}} \approx 200$~K$, T_{\mathrm{PG}}^{\mathrm{V}}\approx 400$~K for \ybcoE. These values are less than what relaxation shows, but sufficiently close for the assumed simple scenario and perhaps inhomogeneous samples (see below). 

An important feature of this pseudogap is a high temperature shift offset. It arises from the fact that even far above the pseudogap energy one still misses the low energy states. Even if the shifts are temperature independent, they can carry a doping dependence (as the pseudogap depends on doping), i.e. two variables are needed to describe the shifts ($K(x,T)$).

\section{Planar Oxygen Relaxation in other Cuprates}
\begin{figure}[h!]
\centering
\includegraphics[width=1\textwidth ]{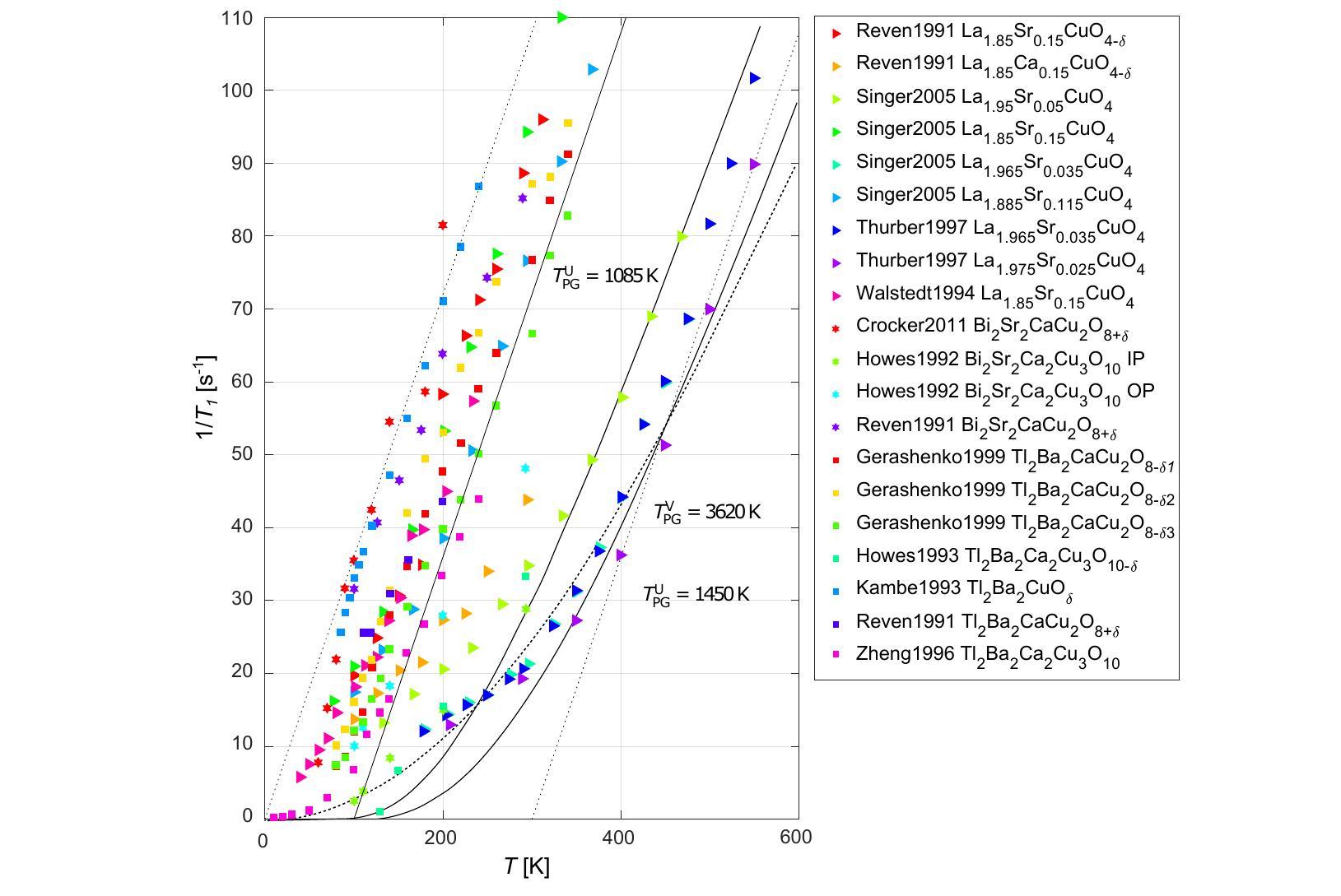}
\caption{Planar O relaxation rates ($c\parallel B_0$) as a function of temperature for other cuprates. The slopes are rather similar to those observed for \ybco and \ybcoE in Fig.~\ref{fig:fig1} as the dotted lines show. The U-shaped gap closes rapidly with increasing doping where all low energy states are recovered. The maximum slope (DOS) appears to be a property of the CuO$_2$ plane, as well as the maximum size of the gap. For the references see Appendix {\bf B}.  
}\label{fig:fig5}
\end{figure}

In Fig.~\ref{fig:fig5} we plot relaxation data from the literature for all other cuprates. Note that only the temperature axis is different (up to \SI{600}{K}) compared to that in Fig.~\ref{fig:fig1}{\bf B, C}.

We note that the slope for optimally and overdoped \ybco (left dashed line) is similar to the dependences found for the other overdoped cuprates. Thus, the CuO$_2$ plane appears to have this upper bound on the DOS. However, if we look at optimally doped \lsco, it appears to still have a sizable pseudogap, in fact similar to that of \ybcoE. The largest gap is observed for the very underdoped \lsco ($x=0.025$) with $T^{\mathrm{U}}_{\mathrm{PG}} \approx 1450$ K, the size of the exchange coupling in the cuprates. A V-shaped gap appears to fit better the low temperature behavior. It could be the states near the gap edge that are special (coherence peaks), also in-gap states could play a role in enhancing the relaxation at low temperature. Again, the loss of parts of the inhomogeneous sample with a large gap favors states from lower gap areas with increased relaxation. Quadrupolar relaxation plays some role, as well. Thus, the shape of the gap cannot be deduced from the low-temperature behavior. The gap rapidly closes with doping, as widely assumed.

Note that the high temperature behavior is similar for all materials, which does support the idea of a temperature independent gap set by doping, and, importantly, very similar high-temperature Fermi liquid-like states.

To conclude, planar O NMR relaxation appears ubiquitous to the cuprates, and it defines and measures the pseudogap in a rather simple way (which is not the case for planar Cu relaxation and shift \cite{Avramovska2019,Jurkutat2019,Avramovska2020}).  

\section{Planar Oxygen Shifts in other Cuprates}
\begin{figure}[h!]
\centering
\includegraphics[width=1\textwidth ]{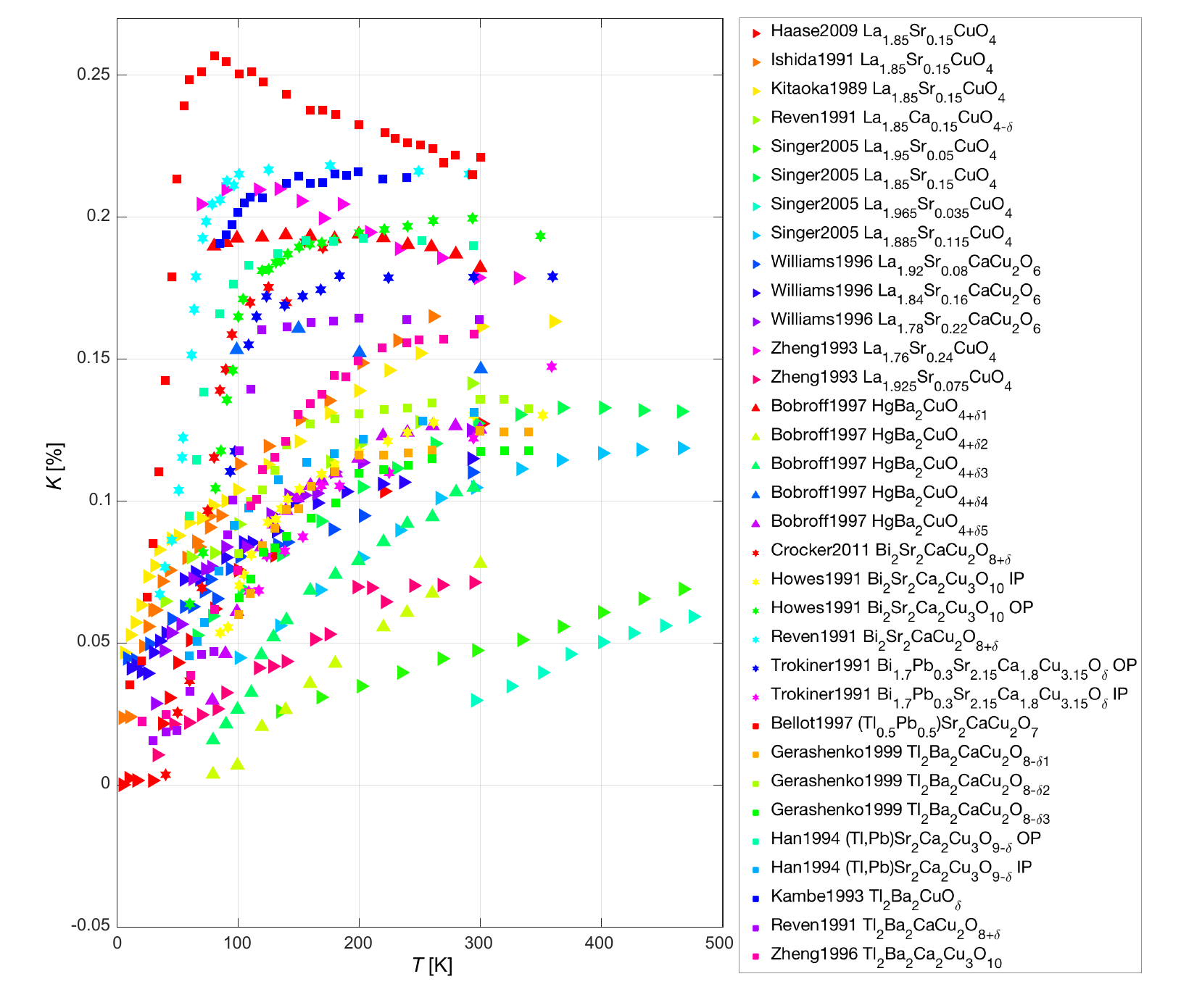}
\caption{Planar $^{17}$O NMR shifts for ${c\parallel B_0}$ for the other cuprates. Note that the temperature axis extends to 500 K. For more detailed plots see Figs.~\ref{fig:fig7} and \ref{fig:fig8}. Note that a high-temperature independent shift may still show lost states, as for optimally doped \lsco. For references see Appendix B.   
}\label{fig:fig6}
\end{figure}

Shift data from all other materials are presented in Fig.~\ref{fig:fig6}. The overall qualitative phenomenology is similar to what was found for \ybco and \ybcoE. 
Except for a couple of overdoped materials, the shifts increase monotonously with temperature. Overdoped systems have nearly temperature independent shifts, as for a Fermi liquid, and drop rapidly near \tc. In the pseudogap regime the shifts begin to show a temperature dependence above \tc, however, a temperature independent shift as for La$_{1.85}$Sr$_{0.15}$CuO$_4$ at high temperatures does not mean there is no pseudogap. Again, Fermi liquid-like shifts can be suppressed in the cuprates due to lost, low-energy states \cite{Avramovska2019}. 

The  superconducting gap is hardly noticeable, as there are no rapid changes of the shifts near \tc. Despite the scarcity of data below \tc, it appears that a number of materials could show a negative shift at the lowest temperatures. 

The maximum observed shifts for overdoped materials are expected from the Korringa ratio by using the dominant slope in the relaxation plots ($1/T_1T \approx $ 0.36/Ks). Samples with the largest pseudogap (La$_{1.965}$Sr$_{0.035}$CuO$_4$) also have the lowest high temperature shifts. Obviously, the pseudogap can lead to doping-dependent, but not necessarily temperature dependent spin shift ($K(x,T)$) since the low-energy states are still missing for small pseudogaps at high temperatures.

The true temperature dependence of the shifts in the pseudogap region is difficult to assess as sample inhomogeneity leads to a loss of the shift from areas that show a larger pseudogap as the temperature is lowered, cf. dashed lines in Fig.~\ref{fig:fig4}. 

It is also clear that optimally doped materials may have almost no pseudogap as for \ybco, but it can be sizable as for \lsco. 


\begin{figure}[h!]
\centering
\includegraphics[width=1\textwidth ]{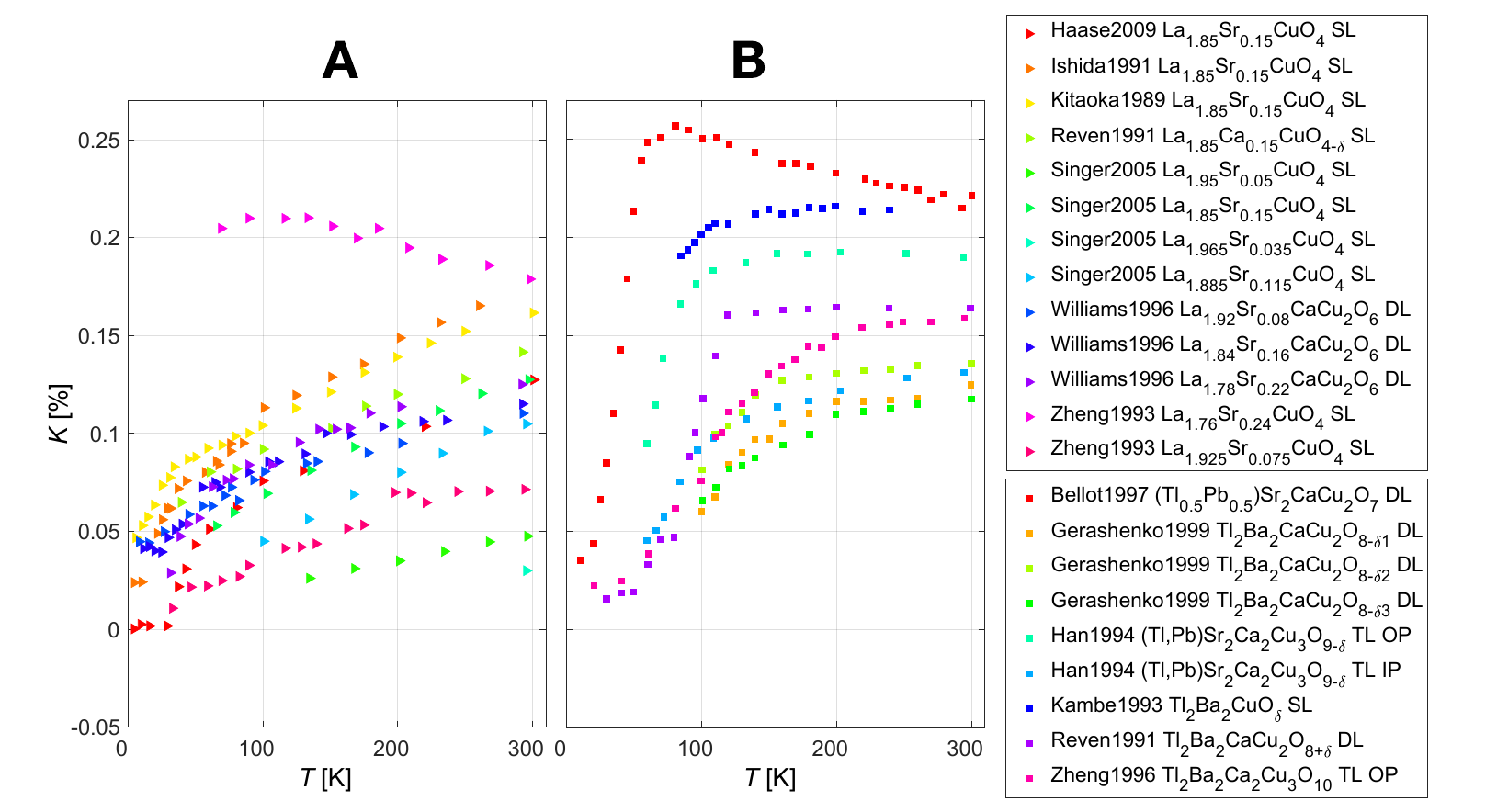}
\caption{Planar $^{17}$O NMR shifts for ${c\parallel B_0}$ for the other cuprates from Fig.~\ref{fig:fig6}, separated for clarity. {\bf A},  La based cuprates, single and double layered. The doping ranges from x=0.035, highly underdoped (lowest point near \SI{300}{K}), to x=0.24, highly overdoped. The shifts cover the range from 0.01\% to 0.2\%. The highly overdoped sample has the highest shift (there is some discrepancy between optimally doped data from different sources, probably due to inhomogeneity). 
{\bf B}, Tl based compounds. The overdoped samples have the highest and Fermi liquid-like shifts and also show an abrupt decrease near $T_{\mathrm{c}}$. As doping is lowered the shifts become more suppressed. In the triple layer compound the inner plane (IP) has a larger pseudogap than the outer plane (OP).
}\label{fig:fig7}
\end{figure}

\section{Discussion and Conclusions}
Planar O relaxation and spin shift data were collected and simple plots reveal that they demand a temperature independent pseudogap at the Fermi surface with a size set by doping. The pseudogap rapidly opens, coming from the overdoped side by decreasing doping, and it approaches the size of the exchange coupling, $J$, for strongly underdoped systems. The states above the pseudogap, no matter what its size is, appear to be the same for all cuprates and carry even a more or less constant density, as perhaps expected from a two-dimensional surface. In fact, in the absence of this pseudogap, shift and relaxation for planar O are Fermi liquid-like and the Korringa relation holds. This supports the view that even in the presence of the pseudogap, the available states above it are the same Fermi liquid-like states. 
The doping level at which the pseudogap disappears can be different for different materials. For example, at optimal doping there is a substantial pseudogap already present for \lsco, while the pseudogap has vanished for optimally doped \ybco. For triple layer materials the pseudogap is much larger for the inner layer. A plot of the pseudogap temperature for a U-shaped gap ($T_\mathrm{PG}^\mathrm{U}$) is shown in Fig.~\ref{fig:fig9}.

An important consequence of the temperature independent pseudogap is a doping dependent spin shift. At high temperatures where the shifts can be nearly temperature independent (Fermi liquid-like), states can still be missing and thus the magnitude of shift can be suppressed. Consequently, the cuprate planar O spin shifts must carry at least two independent variables, one related to doping and the other to temperature, $K(x,T)$. This is effectively a two-component description. Whether this two-component description is sufficient is not clear (for planar Cu it is not \cite{Haase2017}). 

At lower energies, there are deviations from the simple behavior, but it is difficult to analyze given the possible influence of inhomogeneity and quadrupolar relaxation. Likely, states in the gap or near the gap edge are responsible for special behavior. 

Very recently, it was shown from plots of literature shift data of planar Cu \cite{Haase2017} that there is a doping dependent spin shift at high temperatures, and comparison with planar Cu relaxation data \cite{Avramovska2019,Jurkutat2019} - that do not show a pseudogap  - led to the conclusion of suppressed planar Cu spin shifts \cite{Avramovska2019,Avramovska2020}, as well. Thereafter, it was shown that this doping dependent planar Cu spin shift explains the conundrum of the correlation of high-temperature spin shifts with the local charge \cite{Pavicevic2020}, resulting in the hallmark asymmetric total planar O lineshapes (that include the quadrupolar satellites) of the cuprates \cite{Haase2000,Pavicevic2020}. 

\begin{figure}[h!]
\centering
\includegraphics[width=.9\textwidth ]{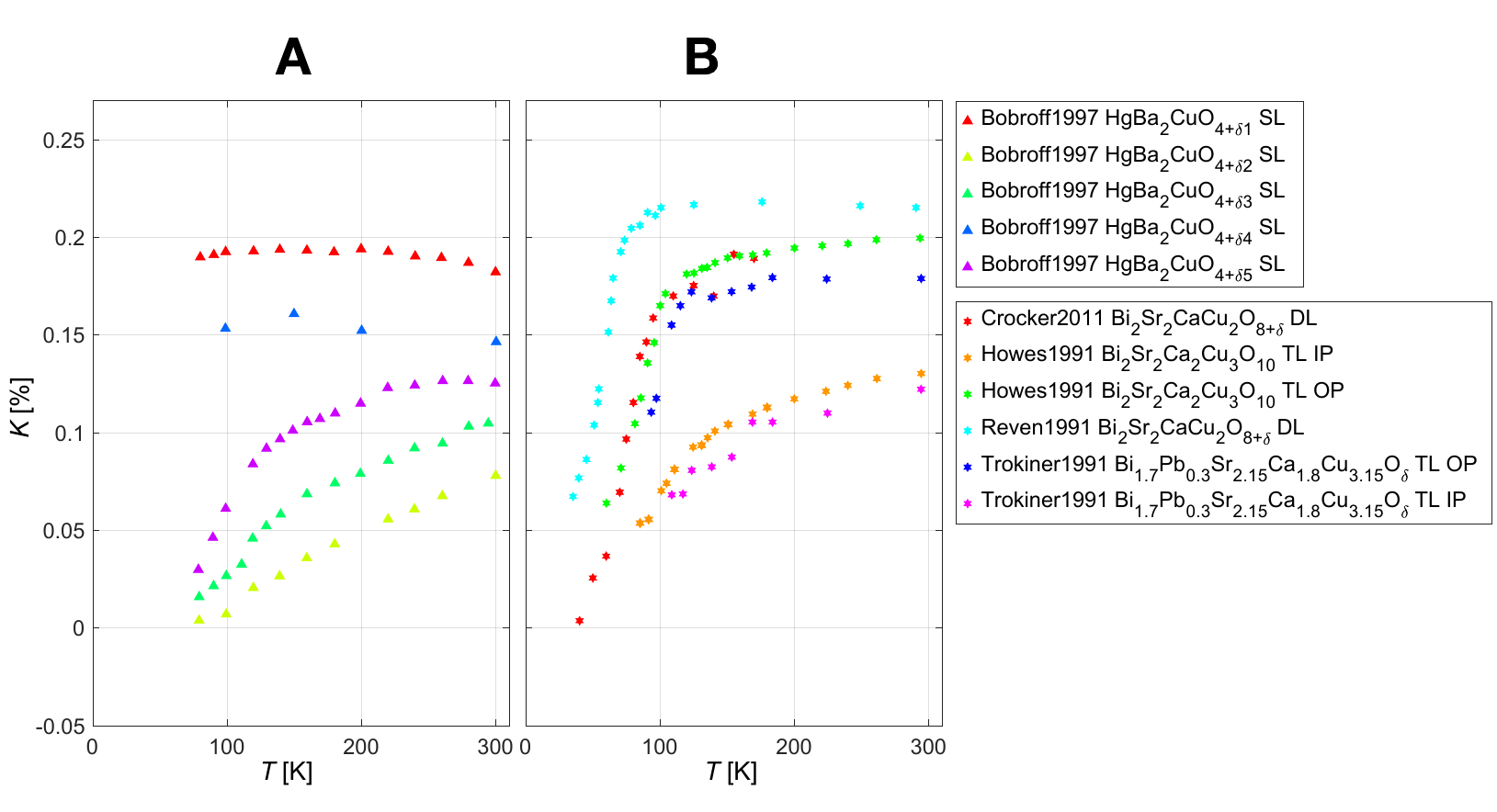}
\caption{Planar $^{17}$O NMR shifts for ${c\parallel B_0}$ for the other cuprates from Fig.~\ref{fig:fig6}, separated for clarity. {\bf A}, single layer mercury based cuprates; the two overdoped samples have temperature independent shifts. The pseudogap becomes apparent at optimal doping (purple triangles). {\bf B}, Bi based cuprates; the two double layered and overdoped samples have the highest and temperature independent shifts, with an abrupt drop near $T_{\mathrm{c}}$. The outer plane shifts from the two triple layered compounds show Fermi liquid-like behavior, whereas the inner plane (yellow and pink stars) show a large pseudogap.   
}\label{fig:fig8}
\end{figure}

Here, we argue that it is the doping dependence of the pseudogap that plays the dominant role for these effects. Then it follows that it is the pseudogap that can be spatially very inhomogeneous \cite{Pavicevic2020}. This distinction could not be made earlier \cite{Haase2000}, but it is in agreement with STM data \cite{Pan2001}. 
With a large distribution of the pseudogap, shift and relaxation can be affected. An inhomogeneous broadening changes the apparent temperature dependence of the shift, cf. Fig.~\ref{fig:fig4}, as small pseudogap areas contribute more to the shift at lower temperatures than those with large pseudogaps. For relaxation, the faster relaxing regions, i.e. those with a smaller pseudogap, may dominate throughout the whole temperature range, if spin diffusion in possible. Thus, one has to be very careful in analyzing shift and relaxation quantitatively \cite{Bussmann2011}. 

The inhomogeneity of the pseudogap affects the apparent temperature dependence of the average shift, as discussed with the dashed lines in Fig.~\ref{fig:fig4}, but also the observed linewidths depend on it. In view of Fig.~\ref{fig:fig4} one concludes that in case of an inhomogeneity of the pseudogap the NMR linewidths grow towards lower temperatures before they finally decrease again, while the shift is decreasing monotonously. This is exactly what was found experimentally (for \ybco and \lsco \cite{Haase2000}), and what was interpreted as proof for two different spin components \cite{Pavicevic2020}. 

The relation of this pseudogap to the intra-unit cell charge variation that was first proposed from NMR data \cite{Haase2003} and very recently shown to exist in the bulk of the material \cite{Reichardt2018} is not clear. However, the response of the local charge symmetry to an external magnetic field and pressure found with NMR \cite{Reichardt2016,Reichardt2018}, must bear similarities to the discussed charge ordering phenomena and special susceptibilities associated with the pseudogap \cite{Mukhopadhyay2019,Sato2017}, recently. The total charge involved in the ordering is small (1-2 \% of the total planar O hole content) and may come from states within the pseudogap.

Note that the superconducting transition temperature \tc appears to be not affected by this inhomogeneity, as it is nearly proportional to the average planar oxygen hole density of the parent compounds \cite{Jurkutat2014,Rybicki2016}. Then, with the size and distribution of the pseudogap set by doping, there appears no simple relation to the maximum \tc. 

\begin{figure}[h!]
\centering
\includegraphics[width=.9\textwidth ]{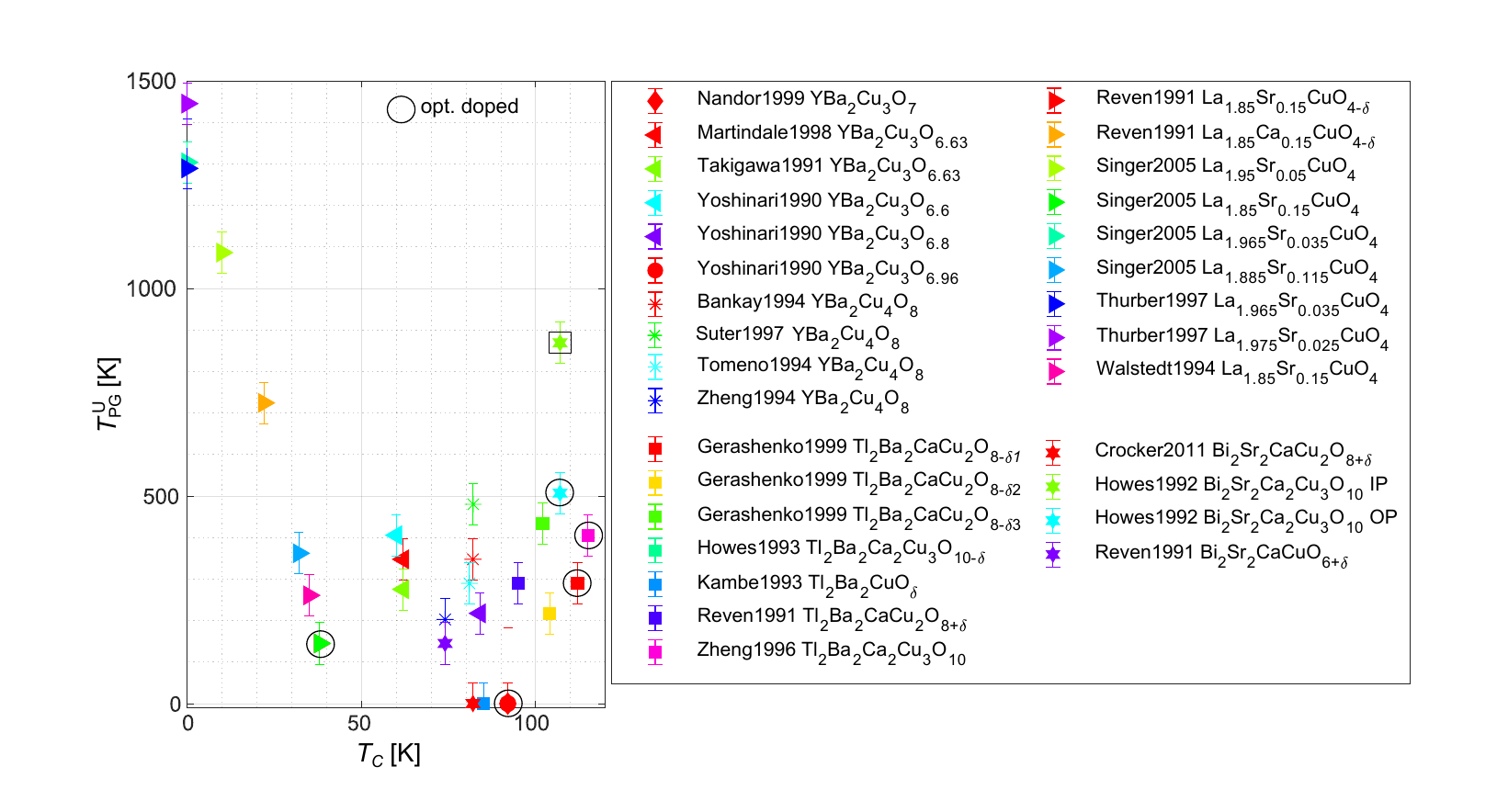}
\caption{Values for a U-shaped gap, $T_\mathrm{PG}^\mathrm{U}$, as determined from the relaxation data vs. the critical temperature, $T_\mathrm{c}$. Optimally doped materials (denoted with circles) can have a vanishing pseudogap as for YBa$_2$Cu$_3$O$_{6.96}$ despite a rather high $T_\mathrm{c}$, but it appears that materials with the highest $T_\mathrm{c}$ all have a substantial pseudogap, and their T$_\mathrm{c}$ increases with the pseudogap temeprature. The inner layer of the triple layer system (denoted with a square) has a significantly larger pseudogap than the outer layers. These findings are in qualitative agreement with the shift data. The data can also be found in the tables in the appendices.   
}\label{fig:fig9}
\end{figure}

The pseudogap behavior was first reported with measurements above \tc for $^{89}$Y NMR of \ybco \cite{Alloul1989}, and  these data show a high-temperature offset in the shifts, as well. So we believe that $^{89}$Y NMR data are in agreement with what we found for planar O here. 

A U-shaped gap in our simulation means that all states contributing to planar O relaxation vanish suddenly within the gap. With such an assumption the largest pseudogap appears to be set by the exchange coupling. Then, effectively, doping decreases the energy gap that needs to be overcome for electrons to flip the nuclear spin for relaxation. Of course, the true shape of the gap and the nature of the states within the gap are not known. 

If the above scenario describes the essential electronic states involved in cuprate conductivity and superconductivity, it should leave its typical signature in electronic specific heat. Indeed, the \ybco family of materials appears to fit the specific heat data by Loram et al. \cite{Loram1998} rather well \cite{Fine2020}. Loram et al. \cite{Loram1998} argue similarly in their specific heat investigations, as the specific heat is linear in temperature in the pseudogap range. Additional states are added by temperature at the same rate as for overdoped systems where there is no gap. Thus, the specific heat of other materials should be similar in view of all analyzed planar O data.

Planar Cu relaxation was shown \emph{not} to be affected by the pseudogap, at all \cite{Avramovska2019, Jurkutat2019}, its relaxation is rather ubiquitous across all cuprates ($1/^{63}T_1T_{\mathrm{c}} \approx$ 21/Ks), independent on doping (the relaxation anisotropy changes with doping \cite{Avramovska2020}). With the cuprate specific heat being in agreement with planar O relaxation, the heat involved with the states that relax planar Cu must be small (perhaps nodal particles). Not surprisingly, the planar Cu shifts, as uniform response, do see the pseudogap. The maximum shift $^{63}K \approx 0.8\%$ is also similar to what follows from the Korringa relation. The details of a comparison between planar Cu and O NMR will be investigated in a forthcoming publication. 

Unfortunately, we feel that it is difficult to conclude on the superconducting gap from the planar O data. An inhomogeneous pseudogap dominates the shifts and the relaxation may be partly electric \cite{Suter2000} in the vicinity of \tc. The latter clearly points to the involvement of charge fluctuations \cite{Reichhardt2004,Mazumdar2018}, very different from the relaxation of planar Cu \cite{Jurkutat2019}, which is also rather ubiquitous at low temperatures in the cuprates, when normalized by \tc  \cite{Jurkutat2019}. Naively, one might assume that the states not already lost to the pseudogap disappear rapidly below \tc, further slowing down relaxation, but the opposite behavior is found, i.e., the rate appears to increase at lower temperature before it finally decreases. This could be due to additional quadrupolar relaxation, alternatively, the magnetic relaxation could show a special increase, but perhaps the inhomogeneity of the pseudogap is most important as regions with fast relaxation (small pseudogap) will dominate.
Details of the spin shift, including the behavior below \tc, are difficult to evaluate, as well, not only due to the inhomogeneity, but also because of the uncertainty of the low-temperature data (loss of signal etc.). A small negative spin shift appears to be observed for a number of materials, which would be expected from the suggested shift scenario \cite{Haase2017,Avramovska2019}. 

To conclude, the planar O data in their entirety reveal a simple temperature independent pseudogap scenario. The gap can be as large as the exchange coupling and vanishes with increasing doping in a family specific way. The states above the pseudogap are unique and Fermi liquid-like for all cuprates and have even constant density. This leads to a relaxation that increases at the same rate with temperature for all cuprates above the pseudogap, and to shifts that become temperature independent. However, depending on the size of the pseudogap (located at lower energies), relaxation and shift can still be suppressed at these higher temperatures. This leads to the otherwise unexpected behavior of shift and relaxation found in NMR. The inhomogeneity of the pseudogap becomes apparent from comparison with the total planar O lineshapes and the planar Cu shifts. No simple relation of the pseudogap to the superconducting transition temperature is found. Note, however, that the planar Cu data do not fit this simple scenario with doping independent relaxation and a two-component shift \cite{Haase2017,Avramovska2019,Jurkutat2019,Avramovska2020}, while similarities exist and need to be explored.

%

\subsection*{Acknowledgements}
We acknowledge the communication with Boris Fine (Moscow), who turned our attention to the specific heat data. We acknowledge support from Leipzig University, and financial support by the German Science Foundation (HA1893-18-1).

\subsubsection*{Author contributions}
J.H. introduced the main concepts and had the project leadership; J.N. led the final literature data collection and its presentation in the manuscript, M.A., D.P., and A.E. were involved in the earlier stage of discussions; R.G., J.N., J.H. worked mainly on the preparation of the manuscript.

\vspace{0.5cm}

\newpage

\section{Appendix A}
List of all references for \ybco and \ybcoE. We found about 36 publications on these materials, out of a total of about 80 papers on all cuprates. If the same data set appears in multiple papers, typically from the same group, we only show the last published account.
\unskip
\begin{table}[h!]
\centering
\caption{\label{App1}References for YBCO literature accounts with critical temperature $T_c$, label as shown in figures, reference link, external magnetic field during measurement and the size of  U-shaped gap from the numerical analysis. All samples were aligned powders, if not stated otherwise$^{*}$.}

\begin{tabular}{llllll}\\

\hline
Compound & $T_c$ & Label & Ref. & Field & $T_{\mathrm{PG}}^{\mathrm{U}}$ \\ 
\hline
YBa$_2$Cu$_4$O$_8$ & 82K & Bankay1994 & \cite{Bankay1994} & 9.03T & 350K \\ 
YBa$_2$Cu$_4$O$_8$ & 82K & Brinkmann1992 & \cite{Brinkmann1992} & & \\ 
YBa$_2$Cu$_4$O$_8$ & 82K & Mangelschots1992 & \cite{Mangelschots1992} & 9.129T & \\ 
YBa$_2$Cu$_4$O$_8$ & 81K & Suter1997 & \cite{Suter1997} & 8.9945T & 490K\\ 
YBa$_2$Cu$_4$O$_8$ & 81K & Tomeno1994 & \cite{Tomeno1994} & 5.71T & 290K\\
YBa$_2$Cu$_4$O$_8$ & 74K & Zheng1992 & \cite{Zheng1992} & 11T & \\ 
YBa$_2$Cu$_4$O$_8$ & 74K & Zheng1993 & \cite{Zheng1993b} & 11T & \\
YBa$_2$Cu$_4$O$_8$ & 74K & Zheng1994 & \cite{Zheng1994} & 4.3/11T & 200K \\ 
\hline
YBa$_2$Cu$_3$O$_{7}$ & 93K & Hammel1989 & \cite{Hammel1989b} & 7.0T & \\ 
YBa$_2$Cu$_3$O$_{7}$ & 92K & Horvatic1989 & \cite{Horvatic1989} & 5.75T & \\ 
YBa$_2$Cu$_3$O$_{6.65}$ & 61K  & Kitaoka1989 & \cite{Kitaoka1989} & 5.75T & \\
YBa$_2$Cu$_3$O$_{7}$ & 92K & Kitaoka1989 & \cite{Kitaoka1989} & 5.75T & \\ 
YBa$_2$Cu$_3$O$_{7}$ & 91.2K & Martindale1993 & \cite{Martindale1993} & 0.67T & \\ 
YBa$_2$Cu$_3$O$_{7}$ & 91.2K & Martindale1993 & \cite{Martindale1993} & 8.30T &\\ 
YBa$_2$Cu$_3$O$_{7}$ & 93K & Martindale1994 & \cite{Martindale1994} & 0.67T & \\ 
YBa$_2$Cu$_3$O$_{7}$ & 93K & Martindale1994 & \cite{Martindale1994} & 8.30T & 0K\\ 
YBa$_2$Cu$_3$O$_{6.63}$ & 62K & Martindale1998 & \cite{Martindale1998} & high field &350K \\ 
YBa$_2$Cu$_3$O$_{6.96}$ & 92.2K & Martindale1998 & \cite{Martindale1998} & high field &  \\
YBa$_2$Cu$_3$O$_{7}$ $^{*}$& 92K & Nandor1999 & \cite{Nandor1999} & 9.05T & 0K\\ 
YBa$_2$Cu$_3$O$_{7}$ $^{*}$& 92K & Reven1991 & \cite{Reven1991} & 8.45T & \\ 
YBa$_2$Cu$_3$O$_{7}$ & 93K & Takigawa1989 & \cite{Takigawa1989b} & & 0K\\
YBa$_2$Cu$_3$O$_{6.63}$ & 62K & Takigawa1991 & \cite{Takigawa1991} & 6/7T & 280K \\ 
YBa$_2$Cu$_3$O$_{6.60}$ & 60K & Yoshinari1990 & \cite{Yoshinari1990} & 10T & 410K \\ 
YBa$_2$Cu$_3$O$_{6.80}$ & 84K & Yoshinari1990 & \cite{Yoshinari1990} & 10T & 220K \\
YBa$_2$Cu$_3$O$_{6.96}$ & 92K & Yoshinari1990 & \cite{Yoshinari1990} & 10T &  0K\\ 
YBa$_2$Cu$_3$O$_{6.96}$ & 87K & Yoshinari1992 & \cite{Yoshinari1992} & 8.97T & 0K \\ 
\end{tabular}
\end{table}
\vspace{-1cm}
\section{Appendix B}

Here we list the references for other cuprates, about 44 publications with relevant data. If a data set appeared in multiple papers, typically from the same group, we only show the last published account.
\vspace{-2cm}
\begin{table}[h!]
\centering
\caption{References to literature accounts of data, with critical temperature $T_{\mathrm{c}}$, label as shown in figures, reference link, sample type [a.p.(c.) - aligned powder (crystal); r.p. - randomly orientated powder; s.c. - single crystal], external magnetic field for measurement, and the U-shaped gap size from the numerical analysis [i.p. - inner plain; o.p. - outer plane in case of triple layer compound].}
\begin{tabular}{llllllll}
\hline
Compound & $T_c$ & Label & Ref. & Sample & Field & $T_{\mathrm{PG}}^{\mathrm{U}}$\\ 
\hline
La$_{1.85}$Sr$_{0.15}$CuO$_4$ &  38K & Haase2009& \cite{Haase2009b} & a.p. & 9T & \\
La$_{1.85}$Sr$_{0.15}$CuO$_4$ & 38K & Ishida1991 & \cite{Ishida1991} & a.p. & 11T &\\ 
La$_{1.85}$Sr$_{0.15}$CuO$_4$ &   & Kitaoka1989 & \cite{Kitaoka1989} & a.p. & 5.75T& \\
La$_{1.85}$Ca$_{0.15}$CuO$_{4+\delta}$ & 22K & Reven1991 & \cite{Reven1991} & a.c & 8.45T& 720K \\ 
La$_{1.85}$Sr$_{0.15}$CuO$_{4+\delta}$ & 38K & Reven1991 & \cite{Reven1991} & a.c & 8.45T & 140K\\ 
La$_{1.95}$Sr$_{0.05}$CuO$_4$ & $\sim$10K  & Singer2005 & \cite{Singer2005} & a.c. & 9T& 1085K \\ 
La$_{1.85}$Sr$_{0.15}$CuO$_4$ & 38K  & Singer2005 & \cite{Singer2005} & a.c. & 9T& 140K \\ 
La$_{1.965}$Sr$_{0.035}$CuO$_4$ & 0K  & Singer2005 & \cite{Singer2005} & a.c. & 9T& 1300K \\ 
La$_{1.885}$Sr$_{0.115}$CuO$_4$ & $\sim$32K  & Singer2005 & \cite{Singer2005} & a.c. & 9T & 360K\\ 
La$_{1.965}$Sr$_{0.035}$CuO$_4$ & 0K  & Thurber1997 & \cite{Thurber1997} & s.c. & 9T& 1290K \\ 
La$_{1.975}$Sr$_{0.025}$CuO$_4$ & 0K  & Thurber1997 & \cite{Thurber1997} & s.c. & 9T& 1450K \\
La$_{1.85}$Sr$_{0.15}$CuO$_{4+\delta}$ & 35K & Walstedt1994 & \cite{Walstedt1994} & a.p. & & 260K \\
La$_{1.92}$Sr$_{0.08}$CaCu$_2$O$_6$ & 17.7K & Williams1996 & \cite{Williams1996} & r.p. & 8.45T & \\
La$_{1.84}$Sr$_{0.16}$CaCu$_2$O$_6$ & 31.5K & Williams1996 & \cite{Williams1996} & r.p. & 8.45T & \\
La$_{1.78}$Sr$_{0.22}$CaCu$_2$O$_6$ & 47K & Williams1996 & \cite{Williams1996} & r.p. & 8.45T &\\
La$_{1.76}$Sr$_{0.24}$CuO$_4$  & 25K  & Zheng1993 & \cite{Zheng1993} & a.c. & &\\ 
La$_{1.925}$Sr$_{0.075}$CuO$_4$  & 20K  & Zheng1993 & \cite{Zheng1993} & a.c. & &\\
\hline
HgBa$_2$CuO$_{4+\delta1}$ & 61K & Bobroff1997 & \cite{Bobroff1997} & a.c. & 7.5T &\\ 
HgBa$_2$CuO$_{4+\delta2}$ & 75K & Bobroff1997 & \cite{Bobroff1997} & a.c. & 7.5T &\\ 
HgBa$_2$CuO$_{4+\delta3}$ & 87.8K & Bobroff1997 & \cite{Bobroff1997} & a.p. & 7.5T &\\ 
HgBa$_2$CuO$_{4+\delta4}$ & 89K & Bobroff1997 & \cite{Bobroff1997} & a.p. & 7.5T &\\ 
HgBa$_2$CuO$_{4+\delta5}$ & 95.7K & Bobroff1997 & \cite{Bobroff1997} & a.p. & 7.5T &\\
\hline
Bi$_2$Sr$_2$CaCu$_2$O$_{8+\delta}$ & 82K & Crocker2011 & \cite{Crocker2011} & a.p. & 9T & 0K\\
Bi$_2$Sr$_2$Ca$_2$Cu$_3$O$_{10}$ & 107K & Howes1991 & \cite{Howes1991} & a.p. & 8.45T  &\\ 
Bi$_2$Sr$_2$Ca$_2$Cu$_3$O$_{10}$ & 107K  & Howes1992 & \cite{Howes1992} & r.p. & 8.45T & i.p. 870K\\
Bi$_2$Sr$_2$Ca$_2$Cu$_3$O$_{10}$ & 107K  & Howes1992 & \cite{Howes1992} & r.p. & 8.45T &  o.p. 510K\\
Bi$_2$Sr$_2$CaCu$_2$O$_{8+\delta}$ & 74K & Reven1991 & \cite{Reven1991} & r.p. & 8.45T & 140K\\ 
Bi$_2$Sr$_2$CaCuO$_{6+\delta}$ & 5.6K & Reven1991 & \cite{Reven1991} & r.p. & 8.45T &\\
Bi$_{1.7}$Pb$_{0.3}$Sr$_{2.15}$Ca$_{1.8}$Cu$_{3.15}$O$_\delta$ & 110K & Trokiner1991 & \cite{Trokiner1991} & r.p. &  & \\

\hline 
(Tl$_{0.5}$Pb$_{0.5}$)Sr$_2$CaCu$_2$O$_7$ & 65K & Bellot1997 & \cite{Bellot1997} & r.p. & 7T &\\
Tl$_2$Ba$_2$CaCu$_2$O$_{8-\delta1}$ & 112K & Gerashenko1999 & \cite{Gerashenko1999} & a.c. &  & 290K\\ 
Tl$_2$Ba$_2$CaCu$_2$O$_{8-\delta2}$ & 104K & Gerashenko1999 & \cite{Gerashenko1999} & a.c. &  & 220K\\ 
Tl$_2$Ba$_2$CaCu$_2$O$_{8-\delta3}$ & 102K & Gerashenko1999 & \cite{Gerashenko1999} & a.c. &  & 430K\\ 
(Tl,Pb)Sr$_2$Ca$_2$Cu$_3$O$_{9-\delta}$ & 124K & Han1994 & \cite{Han1994} & r.p.  & 8.45T &\\ 
Tl$_2$Ba$_2$CuO$_7$ & <4.2K & Kambe1991 & \cite{Kambe1991} & a.c. &  &\\ 
Tl$_2$Ba$_2$CuO$_\delta$ & 85K & Kambe1993 & \cite{Kambe1993} & a.c. & 12T & 0K\\
Tl$_2$Ba$_2$Ca$_2$Cu$_3$O$_{10-\delta}$ & 125K & Howes1993 & \cite{Howes1993} & s.c. & 8.45T & 870K\\ 
Tl$_2$Ba$_2$CaCu$_2$O$_{8+\delta}$ & 95K & Reven1991 & \cite{Reven1991} & r.p. & 8.45T  & 290K\\ 
Tl$_2$Ba$_2$Ca$_2$Cu$_3$O$_{10}$ & 125K & Zheng1995 & \cite{Zheng1995} & a.p. & 11T &\\
Tl$_2$Ba$_2$Ca$_2$Cu$_3$O$_{10}$ & 125K & Zheng1996 & \cite{Zheng1996} & a.p. & 11T & 410K\\
\end{tabular}
\end{table}

\newpage

\section{References}
\bibliography{Oxygen}


\end{document}